\documentstyle[aps,twocolumn,epsf,rotate]{revtex}
\begin{document}
\draft
\twocolumn[\hsize\textwidth\columnwidth\hsize\csname
@twocolumnfalse\endcsname

\title{Spectral Properties of the Attractive Hubbard Model}
\author{M.~Letz, R.~J.~Gooding}
\address{Dept. of Physics, Queen's University, Kingston, Ontario, 
Canada K7L 3N6}
\date{\today}
\maketitle
\begin{abstract}
Deviations from Fermi liquid behavior are well documented in the 
normal state of the cuprate superconductors, and some of these differences
are possibly related to pre-formed pairs appearing at temperatures above T$_c$. 
In order to test these ideas we have investigated the attractive Hubbard
model within a self-consistent, conserving ladder approximation. In this
version of the theory, no feature is present which
can be related to the pseudo gap found in the high-T$_c$ materials.
Further, the interactions between two-particle bound states
change the physics of the superconducting instability in a profound fashion,
and lead to a completely different phenomenology that one predicts based on
the non-self-consistent version of the same theory.

\end{abstract}
\pacs
{PACS: 71.27.+a 
}
]
\narrowtext
   \ \ 
\newpage
   \ \ 
\newpage
Numerous recent experiments ({\em e.g.}, ARPES, optical, neutron scattering,
NMR) have shown that in the high T$_c$ cuprate superconductors 
a so-called pseudo-gap is present \cite {houston}; 
this is further supported by the extensive theoretical work of
Randeria and collaborators \cite{randeriarev}.
 This has led
to proposals that electron pairs form at temperatures well above the 
superconducting transition temperature. 
However, possibly due to phase fluctuations, a
macroscopic phase coherent wave function is not formed, so superconductivity
is not encountered until lower temperatures \cite {emeryhoust}. 

This physics motivates our study of the attractive Hubbard model. The
Hamiltonian for this system is
\begin{equation}
H =-t\sum_{\langle ij \rangle, \sigma} 
( c^\dagger_{i,\sigma} c_{j,\sigma} + {\rm {h.c.}} )
~-~\mid U \mid~\sum_i n_{i,\uparrow} n_{i,\downarrow}
\label{eq:NUHM}
\end{equation}
where the lattice sites of a 2-D square lattice are labeled by $\{i\}$,
the lattice fermion operators are denoted by $c_{i,\sigma}$,
and neighbouring sites are represented by $\langle ij \rangle$.
For any nonzero $\mid U \mid$, two-particle bound states
appear, and are physically related to a pair of electrons lowering
the system's energy when they exist on the same lattice site.
This is certainly the simplest example of a model Hamiltonian which allows 
for one to study the interactions between such electron pairs.

We employ the Brueckner-Hartree-Fock theory (or ladder approximation)
which neglects all crossing diagrams and which is therefore valid only at 
low electronic densities. A preformed pair manifests itself in this theory 
as a two-particle bound state.
In the non-self-consistent version of the theory such a pair of
electrons always has an infinite lifetime. 

By analyzing a fully self-consistent formulation of this theory,
we are able to investigate preformed pairs with arbitrary lifetime. 
That is, if one solves this problem in this approximation,
one includes pair-pair interactions.
Haussmann has argued \cite {haussmann93} (when he investigated a 
three-dimensional continuum model with the same physics) that
the interactions between pairs will be, in first order, of a repulsive nature.
This leads to the dissociation of the pairs causing its finite lifetime.

The equations for the Green's function ($G$), pair susceptibility ($\chi$), 
four-leg vertex function ($\Gamma$), and self energy ($\Sigma$), in a 
conserving approximation are well known:
\begin{eqnarray}
\label{eq:green}
G({\bf k},i \omega _n) &=& \left (G_0({\bf k},i \omega _n)^{-1} - \Sigma({\bf k}
,i
\omega _n) \right )^{-1} \\
\label{eq:chi}
\chi({\bf K},i \Omega _n) &=& \frac{-1}{N \beta} \sum_{m,{\bf k}}  
G({\bf K}-{\bf k},i \Omega _n - i \omega _m) G({\bf k},i \omega _m) \\
\label{eq:gamma}
\Gamma ({\bf K},i \Omega _n) &=& \mid U \mid / 
\left ( 1 + \mid U \mid \; \chi({\bf K},i \Omega _n) \right ) \\
\label{eq:self}
\Sigma({\bf k},i \omega _n) &=&  \frac{1}{N \beta} \sum_{m,{\bf q}}
\Gamma ({\bf k}+{\bf q },i \omega _n + i \omega _n) G({\bf q},i \omega _m)
\end{eqnarray}
where the wave vectors and Matsubara frequencies have their usual meaning.
In a non self-consistent theory, one replaces the full Green's functions $G$
in Eqs.~(3,5) with the noninteracting Green's functions $G_0$.
The usual Thouless criterion begins with the non self-consistent theory
and identifies an instability \cite{mattuck} of the normal state in terms 
of poles of the pair susceptibility (in a weakly interacting 3-D system,
the Thouless criterion becomes identical to the equation which
determines T$_c$ in the BCS theory). The changes in this phenomenology that
arise when one includes pair--pair interactions may be an important
component of the high T$_c$ problem.

For full self consistency, this set of equations has to be solved iteratively.
Since such solutions are difficult and time consuming to obtain, we have investigated 
a simple approximation that allows for extensive numerical investigations of 
the resulting equations. To be specific, during the first step of the 
iteration process leading to self consistency,  we make an approximation for 
the pair susceptibility as being equal to the ${\bf k}$-average (denoted
from now on as over-lined quantities, {\em e.g.} $\overline{\Gamma}$) of
the noninteracting pair susceptibility. We only calculate
the ${\bf k}$-averaged pair susceptibility during subsequent iterations
to self consistency. This leads to the following expressions
\begin{eqnarray}
&\overline{\Gamma}&(i \Omega _n)=\frac{1}{N}\sum_{\bf K}\Gamma
({\bf K},i \Omega _n)\approx \mid U \mid /(1+\mid U \mid\overline{\chi}
(i\Omega_n) ) \\
&{G}&({\bf k},i \omega _n)  \approx \left ({G_0}({\bf k},i \omega
_n)^{-1} - 
\overline{\Sigma}(i
\omega _n) \right )^{-1} 
\end{eqnarray}

Besides the computational simplicity of this approach, one important additional 
advantage of this technique arises from the ease with which low temperatures
may be reached for two and higher dimensional systems. Other methods have great 
difficulty reaching the low temperatures for which superconducting properties 
may become apparent, potentially making our approach quite valuable.

Our ${\bf k}$-averaged approximation becomes accurate in any of the following limits:
Quite trivially, (I) At large temperatures all correlations are thermally 
washed out, and for small bandwidths one obtains the atomic limit. In
both cases, ${\bf k}$-dispersion of the pair susceptibility is irrelevant. 
(II) For large U, $\Gamma$ is basically determined by the pole of the bound state 
at $\chi = 1/U$ which means that a weakly dispersive two particle bound
state is well separated from the continuum. So, in this case the
average over ${\bf k}$-space is a good approximation, since the bound state
below the continuum is now replaced by it's ${\bf k}$-average. 
(III) Probably the most interesting limit in which our approximation is valid is
that of large spatial dimensions. It was argued by Vollhardt and collaborators
\cite{vollhardt90} that for a system with large dimensions the ${\bf k}$-dispersion 
of the self-energy vanishes. Such an approximation gives several results which
are also valid in lower dimensions. This limit gives us
a tool to compare our final result with well known results
obtained from the Kondo-impurity problem. This is due to the
possibility of mapping the attractive Hubbard model, via a particle-hole
transformation, onto a positive U Hubbard model in a magnetic
field. The repulsive Hubbard model, in the limit of large
spatial dimensions, becomes equivalent to the Kondo impurity 
problem \cite{tschork}.
We will discuss more fully the regimes of validity of this 
approximation in a future publication.

\begin{figure}
\unitlength1cm
\epsfxsize=12cm
\begin{picture}(7,7.5)
\put(-2.2,-1.3){\rotate[r]{\epsffile{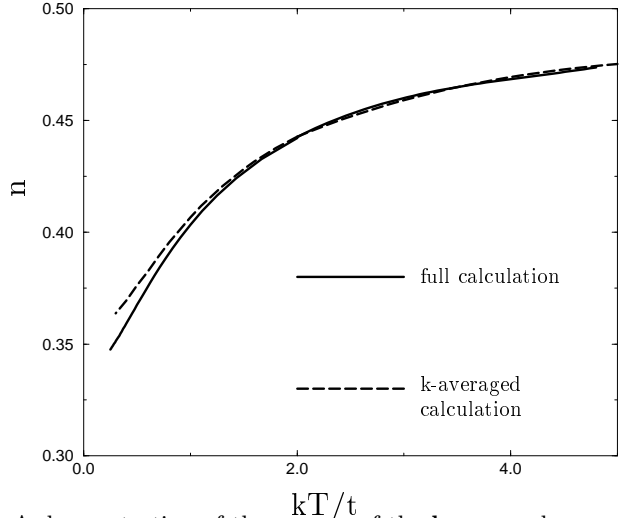}}}
\end{picture}
\caption{A demonstration of the success of the ${\bf k}$-averaged method 
for a one-dimensional system.
Here, $\mid U \mid/t = 4$, namely the Hubbard energy is equal to the
bandwidth, and the chemical potential is fixed to be $\mu/t = -1$. Then,
the electron density n is calculated as a function of temperature. The
solid line shows the ``exact" calculation for the self-consistent, conserving
approximation, and the ${\bf k}$-averaged result is shown by the dashed line.
All energies are given in units of $t$.
\label{fig1}
}
\end{figure}

We begin by showing that for a one dimensional system the ${\bf k}$-average 
method correctly reproduces the physics found in a fully self-consistent,
conserving theory even in the intermediate coupling regime. 
In Fig. \ref{fig1} we have plotted the variation of
electron density with temperature for $\mid U \mid$ being equal to the bandwith
for $\mu = -t$ (note that the free electron band extends from $-2t$ to $2t$).
This quantity was calculated firstly for the self-consistent, conserving theory 
obtained from the ${\bf k}$-average method, and then from the full theory where
${\bf k}$-dispersion was taken into account. The latter work involved
150 points along the imaginary frequency axis and 40 real-space lattice 
points --- we found that these numbers gave fully converged results. 
In both cases the electron density decreases monotonically with
temperature, and there is no sign of a divergence (or strong increase) of the
particle number.

We note that this behaviour is in direct contrast to 
results obtained from non self-consistent calculations. In 
one and two dimensions, for low temperatures, the system is always
unstable towards a condensation into an infinite lifetime
two-particle bound state. For two dimensions this was demonstrated
for a non self-consistent, non conserving theory by
Schmitt-Rink, {\em et al.} \cite{svr}, and we \cite {frankm98} have 
recently shown similar physics is encountered when a fully conserving
theory is used. However, the self-consistent theories, at least to
the minimum temperatures that we have been able to access,
do not show similar physics.

\begin{figure}
\unitlength1cm
\epsfxsize=12cm
\begin{picture}(7,7.5)
\put(-2.2,-1.3){\rotate[r]{\epsffile{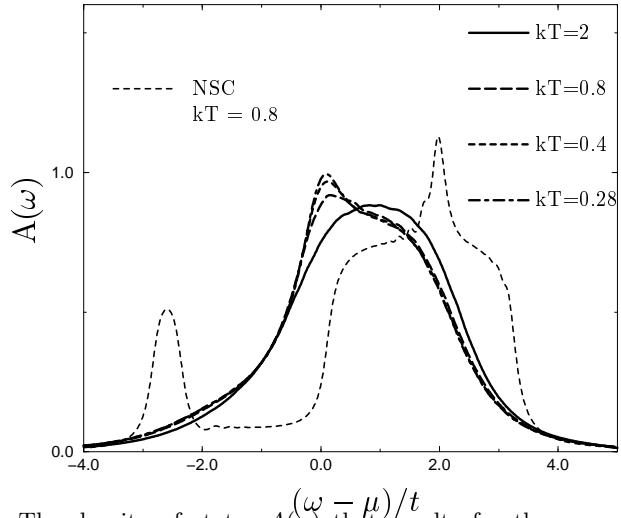}}}
\end{picture}
\caption{The density of states $A(\omega)$ that results for the 
two-dimensional system with n=0.3 and $\mid U \mid/t = 8$ 
for four different temperatures.
Note that the density of states develops a maximum at the
chemical potential when the temperature is decreased. 
\label{fig2}
}
\end{figure}

In this brief report we want to focus on only one aspect of the numerical
results which we have generated from our ${\bf k}$-averaged treatment of
the self-consistent, conserving treatment of the attractive Hubbard model.
In Fig. \ref{fig2} we show the density of states A($\omega$) at
different temperatures  
for a two-dimensional square lattice, again taking $\mid U \mid$ to be equal 
to the bandwidth, at a low electron density $n = 0.3$ ($n = 0.5$ corresponds 
to half filling). We have also shown the results that follow from other 
approximations.

The non self-consistent conserving approximation leads to a
large fraction of the density of states being shifted to a peak below the one
particle continuum --- this peak is due to the infinite lifetime two-particle 
bound state. Between this peak and the one
particle continuum lies a large gap, and this gap has been suggested by some 
to be the origin of the pseudo gap.

However, our self-consistent, conserving, ${\bf k}$-averaged calculation
shows that the gap mentioned above disappears entirely, and that the only 
remnant from the two particle bound state is a flat, strongly lifetime broadened
shoulder below the continuum. Further, instead of a gap, the density of
states becomes a small maximum at the chemical potential, and with
decreasing temperature this maximum increases. The existence of such a
maximum can be understood via the analogy of the attractive Hubbard
model with the repulsive Hubbard model which in large spatial
dimensions can be mapped onto a Kondo impurity model as mentioned
above. In such problems a local maximum of the one particle
density of states also occurs, as seen, {\em e.g.}, by Georges {\em et al.}
\cite{georges93}, who investigated the repulsive Hubbard model by
using an iterated perturbation theory. 

We note that recent work for a $d$-wave pairing interaction \cite{pseudonaz} 
has found similar results, but has been interpreted quite differently.

In conclusion, we have demonstrated the possibility of solving the
attractive Hubbard model in two dimensions in the self-consistent, conserving
ladder approximation by performing a numerically advantageous 
${\bf k}$-average. Due to self-consistency,
we include pair-pair interactions which hinder the formation of an
infinite lifetime bound state. In this way, the Thouless criterion is
altered. Further, evidence for a pseudo-gap in the density of states
is not obvious.

We thank Frank Marsiglio for helpful comments.
This work was supported by the DFG (Deutsche Forschungsgemeinschaft),
and the NSERC of Canada.


\newpage

%
%

\end{document}